\documentstyle[preprint,version2,aps]{revtex}
\begin{document}
\draft
\columnsep -.375in
\begin{title}
Spin Stiffness of Mesoscopic Quantum Antiferromagnets
\end{title}
\author{Daniel Loss $^{1}$ and Dmitrii  L. Maslov $^{1,2}$ \cite{Maslov}}
\begin{instit}

$^{1}$ Department of Physics, Simon Fraser University,
Burnaby B.C. V5A 1S6, Canada

$^{2}$ Department of Physics and Materials  Research Laboratory,
University of Illinois at Urbana-Champaign, IL 61801

\end{instit}

\begin{abstract}
We study the spin stiffness of a  one-dimensional
quantum antiferromagnet in the whole range of system sizes $L$
and temperatures $T$.
We show that  for integer and half-odd integer spin case the stiffness
differs fundamentally in its $L$ and $T$ dependence, and that in the
latter case the stiffness exhibits a striking
dependence on the parity of the number of sites.
Integer spin chains are treated in terms of the  non-linear sigma
model, while half-odd integer spin chains are discussed in
a renormalization group approach leading to
a Luttinger liquid with Aharonov-Bohm type boundary conditions.
\end{abstract}

      Quantum one-dimensional antiferromagnets  have been
the subject of intensive studies since Haldane \cite{Haldane}
conjectured that the spectrum of an integer spin $S$ chain
has a finite gap even in the absence of any anisotropy,
while half--odd integer $S$ chains are gapless. In both
cases the  N\'eel long-range order of the ground state
is destroyed by quantum
fluctuations. However, the \lq\lq\ degree of destruction \rq\rq\
is different: for integer $S$ the correlation length is finite,
which means that the elementary excitations have a gap,
while for half-odd integer $S$ the correlation length is infinite
and excitations are gapless. By now, the
presence of the Haldane gap for integer $S$ chains is well understood
theoretically \cite{Affleck} and has been confirmed in
experimental \cite{exper} and numerical \cite{numer} studies.
However, all these investigations were concentrated on the Haldane gap, i.e.,
on the energy spectrum itself. Thus a broader
understanding of the Haldane conjecture is desirable, and
the question naturally arises whether there are alternative manifestations
of the fundamental difference between integer and half-odd integer spin
chains.
Indeed, it is the purpose of
this work to provide an affirmative answer to this question and to discuss
such a particular case in terms of the so-called spin stiffness.

      Quite recently, there has been much interest in the
spin stiffness (helicity modulus) $\rho _s$ of
{\it classical} Heisenberg ferromagnets
\cite{Chak,Zinn,Caffarel}.
$\rho _s$ is defined as a change in the free energy F of the magnet
when a twist is applied to  the spins  at the sample boundaries.
When thermal fluctuations are taken
into account, $\rho_s$ is being renormalized with respect to its bare value
and depends on the
scale at which it is probed. Chakravarty
\cite{Chak} has recently shown that $\rho_s$ exhibits
features which are familiar from the behavior of
the electrical conductance of a metal in the weak localization
regime
\cite{Alt}. For instance, in $2$d, the mean value of $\rho _s$
depends
logarithmically on the sample size $L$, while the rms value of its
fluctuations is universal\cite{comm1}.
This similarity \cite{footnote2}
makes the
spin stiffness  an equal member of the family of traditionally
mesoscopic quantities such as a conductance or
a persistent current.
Also,  the stiffness $\rho_s$  serves
as  a useful (though not perfect) tool for characterizing magnetic
long-range order; in particular, a vanishing value of $\rho_s$ indicates
absence of order \cite{Goldenfeld}.

     Our goal here is to study $\rho _s$ of
a {\em quantum} one-dimensional antiferromagnet, where fluctuations
are i) quantum and ii) topologically distinct
for integer and half-odd integer $S$.
We shall see  that the behavior of $\rho _s$ is, indeed, quite different
for integer and half-odd integer $S$: $\rho _s$ is  renormalized
with $L$ in the former case (as it is for a classical $2$d ferromagnet),
whereas
it is $L$-independent (in leading order) in the latter.
Moreover, for half-odd integer $S$, $\rho_s$ is shown to exhibit a striking
dependence on the parity of the total number $N$ of spins.
These results
should be amenable to a direct check in numerical simulations (see e.g. Refs.
\cite{Caffarel,Bedell}), and,
in particular,
could be tested experimentally by measuring the stiffness of
quasi--one--dimensional antiferromagnets of finite size using similar materials
as in Ref.\cite{exper}.

      We start with the  Heisenberg Hamiltonian for a spin chain with
nearest-neighbor interactions
 \begin{equation}
 H=J_{ex}\sum ^{N}_{n=1}{\bf S}(n){\bf S}(n+1),
 \label{eq:ham}\end{equation}
 where $J_{ex}>0$, and we consider the integer $S$ case first.
 In this case, the long-wavelength limit of the partition function $Z$
 becomes the $(1+1)$d
 non-linear $\sigma$ model (NL$\sigma$M) \cite{Fradkin}
 with  $Z=\int {\cal D}{\bf n}\delta({\bf n}^2-1)\exp(-{\cal A})$, and
the Euclidean action is given by
 \begin{equation}
 \label{eq:nlsact}
 {\cal A}=\frac{1}{2g}\int ^{L_T}_0 \int ^{L}_0 dx_0 dx
 (\partial _\mu {\bf n})^2,\; \mu=0,1,
 \end{equation}
 where $g=2/S$, $v_s=2SJ_{ex}a_0$ is the spin wave velocity (we set
 $k_B=\hbar=1$), $a_0$ is the lattice constant, $L=a_0 N$,
 $L_T=\beta v_s$ is the wavelength of
 the thermal magnons, and ${\bf n}$ is
 the slow-varying component of the (staggered) magnetization satisfying the
 constraint ${\bf n}^2=1$. On the edges of the
 space--time domain $L\times L_T$
the boundary conditions  are periodic in $x_0$, i.e.,
 ${\bf n}(0, x)={\bf n}(L_T, x)$, and correspond to a fixed twist
 of the ${\bf n}$-field applied in $x$-direction, i.e.,
${\bf n}(x_0, 0)=(1,0,0)$ and
 ${\bf n}(x_0, L)=(\cos \theta, \sin \theta, 0)$, where $\theta$ is the
 twist angle. It is convenient to use the transformation \cite{Chak,Zinn,Zinnb}
 ${\bf n}={\cal R}(\theta(x)){\bf
 \sigma}$, where ${\cal R}$ is the  rotation matrix about
 the $z$-axis  by the angle  $\theta(x)=\theta x/L$, and
 ${\bf \sigma}$
 satisfies the boundary conditions $\sigma_1=1$, and
 $\sigma_{2,3}=0$, at $x=0, L$.
 The action then takes the form
 \begin{eqnarray}
  {\cal A}=\frac{1}{2g}\int
 d{\bf x}[(\partial_\mu {\bf \sigma})^2+\frac{\theta^2}{L^2}
 (1-\sigma_3^2)\nonumber\\
 +2\frac{\theta}{L}\left(\sigma_1\partial
 _x\sigma_2-\sigma_2\partial _x \sigma_1\right)].
 \label{eq:action}\end{eqnarray}

  We  define the spin stiffness $\rho_s$ in units of the velocity
 \begin{equation}
 \label{eq:stiff.def}
 \rho_s=\frac{1}{2}L\frac{\partial ^2F}{\partial \theta^2}\left|_{\theta
=0}\right.,
 \end{equation}
 where $F=-T\ln Z$ \cite{comm.def.}. With this
 definition, the bare value of $\rho_s$ in the tree (classical)
 approximation is $\rho_s^0=v_s/2g$. Corrections due to quantum and thermal
 fluctuations can be found  in a loop expansion in $g$.
 In one-loop order, only quadratic terms in the action (\ref{eq:action})
 are to be retained, and the third term in Eq.~(\ref{eq:action}) reduces
to a total derivative $\partial_x \sigma_2$, and, thus, vanishes
 as we shall restrict our consideration to the topological sector with
 zero winding number (Pontryagin index). Performing the functional
 integration, we get
 \begin{equation}
 \label{eq:stiff.interm}
 \rho_s =\rho_s^0\left(1-\frac{g}{LL_T}\sum_{\bf q}
 \frac{1}{{\bf q}^2}\right),
 \end{equation}
 where ${\bf{q}}~=~(2\pi n/L_T,~\pi m/L)$, $n=0,\pm 1,...$,
 ~$m=1,2$...\nolinebreak
 The sum in Eq.~(\ref{eq:stiff.interm})
 can be evaluated in three limiting cases:
 a) $L\ll L_T$ (quantum region); b) $a_0\ll L_T\ll L$ (classical renormalized
 region); and c) $L_T\ll a_0$ (classical region) .
 We have
 \begin{eqnarray}
 &&\frac{\rho_s}{v_s}=\left\{
 \begin{array}{l}
 \ln\left(\xi_{qm}/L\right)/4\pi,\;\; \mbox{for a)}\\
 (\xi(T)-L)/12L_T,\;\; \mbox{for b)}\\
 (\xi_{cl}-L)/12L_T,\;\; \mbox{for c)}\\
 \end{array}\right.
 \label{eq:rhos.int}
 \end{eqnarray}
Here, $\xi_{qm}=\alpha a_0\exp(\pi S)$ is the correlation
 length in the quantum region.
 $\xi(T)=3 L_T\ln(\gamma\xi_{qm}/L_T)/\pi$ is the classical correlation
 length $\xi_{cl}=6L_T/g$ renormalized by quantum fluctuations.
$\alpha$ and $\gamma$
are (cut-off dependent) non-universal constants of order one.
We note that  case c) agrees with  the $1$d classical
NL$\sigma$M \cite{Chak}.

 In all the regions,  $\rho_s$
 goes to zero as the system size $L<\xi$ approaches  the correlation
 length of the corresponding region \cite{2d.comm}.
 This zero value of $\rho_s$ is what a macroscopic system is expected
to have in the absence of spontaneously broken symmetry
\cite{Goldenfeld} (the point $\rho_s=0$ signals the breakdown
 of  the one-loop order expansion).

 The rms
 fluctuation $\delta \rho_s^2=~(TL
 \partial/\partial \theta^2)^2 ~\ln Z|_{\theta=0}$ is given by
 \begin{eqnarray}
 \label{eq:fluct.int}
 \frac{\delta \rho_s^2}{v_s^2}=\frac{1}{2L^2L^2_T}\sum _{{\bf q}}
\frac{1}{{\bf q}^4}=
 \left\{
  \begin{array}{l}
  \frac{\zeta(3)}{8\pi^3}\frac{L}{L_T},\;\;\mbox{for a)}\\
  \frac{1}{180}\left(\frac{L}{L_T}\right)^2,\;\;\mbox{for b) and c)},\\
  \end{array}\right.
  \end{eqnarray}
where $\zeta(x)$ is the Riemann $\zeta$-function.
Contrary to the case of a classical ferromagnet
\cite{Chak}, the
fluctuations are {\em non-universal}: they depend both on $L$ and $T$.
Moreover, in the classical and classical renormalized regions, the
fluctuations are abnormally large ($\delta \rho_s > \rho_s$), and, thus,
the spin stiffness is not a self-averaging quantity. Finally, we note that
the analogues of the regions we consider in the quantum NL$\sigma$M can
also be obtained
in the classical $2$d model, if one considers a rectangular
instead of a square system.

     We now turn to the half-odd integer spin case. The effective
field-theoretical description of the long-wavelength excitations is not of
much use here, since the partition function contains a $\Theta$--term
and contributions from
all topological sectors with different winding numbers
\cite{Haldane,Affleck,Fradkin}, which makes the model hardly tractable.
It is believed that
the exactly solvable case of the spin 1/2 chain reproduces
the  generic features of all half-odd integer $S$ chains
\cite{Haldane,Affleck,Fradkin}, and we shall consider this case only.
By using the
Jordan--Wigner transformation \cite{JW}
\begin{equation}
\label{eq:jwt}
\psi(n)=(-1)^n\exp\left(i\pi\sum^{n-1}_{j=1}(S_z(j)+\frac{1}{2})\right)S_{-}(n),
\end{equation}
where $S_{\pm }=S_x\pm iS_y$, the  Hamiltonian (\ref{eq:ham}) is
mapped on to a system of spinless  fermions on the lattice,
\begin{eqnarray}
\label{eq:ham.ferm}
H=J_{ex}\sum_{n=1}^{N}(-\frac{1}{2}\left\{\psi^{\dagger}(n)\psi(n+1)
+\mbox{H. c.}\right\}\nonumber\\
+:\rho(n)::\rho(n+1):),
\end{eqnarray}
where $:\rho(n):=\psi^{\dagger}(n)\psi(n)-1/2$ is the (Fermi-ordered)
density operator. We now have to specify the boundary condition for the
fermionic operators $\psi$.
The quantum generalization of the classical
boundary conditions for the spin field ${\bf n}$,  used in the NL$\sigma$M
treatment of the integer $S$ case, is $S_\pm(N+1)=e^{\pm i\theta}S_\pm(1)$, and
$S_z(N+1)=S_z(1)$.
The boundary
condition for $\psi$ then follows from Eq.~(\ref{eq:jwt}),
\begin{equation}
\psi(N+1)~=~e^{i\left[\pi(N_F+N)-\theta\right]}\psi(1),
\label{eq:bcpsi}\end{equation}
 where the number of fermions is
$N_F=N/2$, if $N$ is even, and $N_F=(N+1)/2$, if $N$ is odd. We have also used
the fact that $\sum S_z(n)=0 (1/2)$ for even (odd) $N$.
The problem defined by Eqs.~(\ref{eq:ham.ferm}) and (\ref{eq:bcpsi})
is similar to that of spinless electrons on a ring threaded by an Aharonov-Bohm
flux $\theta$, with the difference that here the boundary conditions depend on
the parity of $N$. This parity dependence
will result in a striking difference in the behavior of
$\rho_s$ for even and odd $N$.

 Finite-size systems of interacting fermions with twisted
boundary conditions have recently been
 studied in the framework of the Luttinger liquid approach \cite{Loss}.
 The parity dependence of the boundary conditions, however, requires
a re-examination
 of the bosonization scheme, which we now address.
The left- and right--movers
are introduced by $\psi(n)=e^{ink_F}\psi_+(n)+e^{-ink_F}\psi_-(n)$,
where we choose $k_F=\pi/2$ for $N$ even and odd.
The boundary
conditions for $\psi_{\alpha}$, where $\alpha=\pm$, take the form
\begin{eqnarray}
\label{eq:bc.ferm}
\psi_{\alpha}(N+1)=
\left\{
\begin{array}{l}
e^{-i\theta}\psi_{\alpha}(1), \;\;\mbox{for}\;N\;\mbox{even}\\
-\alpha i e^{-i\theta}\psi_{\alpha}(1),\;\;\mbox{for}\;N\;\mbox{odd}.\\
\end{array}\right.
\label{eq:bcpsipm}\end{eqnarray}
Bosonic fields are introduced by
$\psi_\alpha=~(2\pi a_0)^{-1/2}~e^{i\sqrt{\pi}\phi_\alpha}$,
where $\phi_\alpha~=\alpha\phi-\vartheta$, and $\partial_x \vartheta$ is the
conjugate momentum of $\phi$.
The zero
modes of  $\phi$ and $\vartheta$ can be chosen in the form \cite{Loss}
\begin{eqnarray}
\label{eq:boson}
&\phi_0=\phi_J/\sqrt{\pi}+{\bf M}\sqrt{\pi}x/L\nonumber,\\
&\vartheta_0=\vartheta_M/\sqrt{\pi}+({\bf J}-
\theta/\pi)\sqrt{\pi}(x+\frac{1}{2}L)/L,
\end{eqnarray}
where  ${\bf J}$ and ${\bf M}$ are the operators
of the topological current and the number of particles above the
ground state \cite{Haldane81}, respectively,
which satisfy
$[\phi_J,{\bf J}]=[\vartheta_M,{\bf M}]=i$.
Next, using the Baker-Hausdorff formula, we write
\begin{equation}
e^{i\sqrt{\pi}\phi_{\alpha}}={\bar\psi} e^{i\frac{\pi}{L}\left[(\alpha x
(M-1)-J(x+\frac{L}{2})\right]},
\label{eq:sep}\end{equation}
where ${\bar\psi}$ contains contributions from the nonzero modes and from
$\phi_J$ and $\vartheta_M$, and is not parity dependent; $J(M)$ stands for
the eigenvalue of ${\bf J}({\bf M})$. It is convenient
to introduce the topological indices $\kappa _J$ and
$\kappa_M$, such that $\kappa_J=0(1)$, if  $J$ is even (odd);
for even $N$,
$\kappa_M=0(1)$, if $M$ is even (odd), and, for odd $N$,
$\kappa_M=0(1)$, if $M+1/2$ is even (odd).
By using Eqs.~(\ref{eq:bcpsipm}) and (\ref{eq:sep}),
we see that  $\kappa_{J,M}$ must satisfy the following constraints:
$\kappa_J=1$, $\kappa_M=0$ (and vice versa), if $N$ is even, and
$\kappa_J=\kappa_M$, if $N$ is odd. By comparing these constraints
 with the analogous
constraints of the fermionic problem \cite{Loss},
we can now say that the ground state of our spin
system is {\em paramagnetic} for $N$ even and {\em diamagnetic} for $N$
odd.

 The  rest of the bosonization procedure is identical to that of
 Ref.~\cite{Loss}, and the bosonized
 Euclidean action takes the sine-Gordon form
\begin{eqnarray}
\label{eq:act.ferm}
&{\cal A}_b=\int_0^{\beta v_0}\int_0^L d^2x\{K_0(\partial _\mu\phi)^2
+(i/L)\sqrt{\pi}\theta_0\partial _0\phi\nonumber\\
&-(g_0/a_0^2)\cos(4\sqrt{\pi}\phi)\},
\end{eqnarray}
where $v_0=v_s(1+4/\pi)^{1/2}$, $K_0=v_0/2v_s$, $g_0=1/8\pi^2 K_0$,
and $\theta_0=\kappa_J-\theta/\pi$. The bosonic fields have been decompactified
in the course of the functional integral derivation, and  $\phi$ obeys
now  the boundary
conditions $\phi(x_0+k_0\beta v_0, x+k_1L)=\phi(x_0,x)+k_0\sqrt{\pi}n+
k_1\sqrt{\pi}(2m+\kappa_M)$, where $n$ and $m$ are the winding numbers
in  $x_0$- and $x$-directions, respectively.  The measure ${\cal D}\phi$
of the functional integral $Z=\int{\cal D}\phi \exp(-{\cal A}_b)$
includes the sums over the winding numbers $n,m$ and  over the topological
indices $\kappa_{J,M}$.
The last term in Eq.~(\ref{eq:act.ferm}) corresponds to Umklapp
scattering processes between fermions.
 Since
 $g_0/K_0$ is small ($\approx0.02$),
 this Umklapp
term can be treated perturbatively in a standard
renormalization group (RG) approach leading to the following flow equations
\cite{KT}:
  \begin{eqnarray}
  &\frac{dg}{dl}=2(K-1)g,\;\;\; g(0)=g_0 \nonumber\\
  &\frac{dK}{dl}=2\pi^2g^2,\;\;\;K(0)=K_0,
  \label{eq:rg}\end{eqnarray}
  where $l=\ln({\cal L}/a_0)$ with  ${\cal L}=\mbox{min}\{L, L_T\}$.
  Since we started with the isotropic Heisenberg model (\ref{eq:ham}),
  the scaling dimension
  of the Umklapp term is equal to the critical dimension of the
  model ($=2$), i.e., this term is marginally relevant. In this case,
  the flow proceeds along the separatrix between massive and
  massless phase.
 On this line \cite{footnote3}, the solutions to
   Eqs.~(\ref{eq:rg}) are with ${\cal L}>> L_0$
   \begin{equation}
   K=K^{*}-\frac{1}{2\ln\frac{{\cal L}}{L_0}},\;\;\;
g=\frac{K^{*}-K}{\pi},
   \label{eq:marg}\end{equation}
where $K^{*}=K(\infty)$ is the fixed point value, and
the (non-universal) cut-off $L_0$ depends on $(K_0, g_0)$ and is of the
order $a_0$.
   At the fixed point (${\cal L}\rightarrow \infty$),
   $g=0$ and the action (\ref{eq:act.ferm}) renormalizes
  to that of a
Luttinger liquid with a topological term ($\propto \partial_0\phi$), and with
parameters  renormalized through  interactions: $(K_0, v_0)
\rightarrow (K^{*}, v^{*})$.  By comparing with
the exact Bethe-ansatz solution \cite{Shankar}, one gets $K^{*}=1$ and
$v^{*}=\pi v_s/2$.

We can  now calculate the fixed-point value of the
spin stiffness, $\rho_s^*$, and its finite size and finite temperature
corrections.
Upon integrating out the zero modes, the twist-dependent part of $Z$ becomes
\begin{equation}
\label{eq:z.ferm}
Z_\theta=\sum_{\kappa_J, \kappa_M}e^{-\kappa_M
b}\theta_3\left(z_J,e^{-a}\right)
\theta_3\left(z_M,e^{-4b}\right),
\end{equation}
where $\theta_3(z,q)=\sum q^{n^2}e^{2inz}$ is the Jacobi $\theta$
function, $z_J=\pi \theta_0/2$,
$a=\pi K^{*}L/L_T^{*}$, $b=\pi K^{*}L_T^{*}/L$, $z_M=2i\kappa_M b$, and
$L_T^{*}=\beta v^{*}$.
The results for the spin stiffness  take
simple forms in the limiting cases of low and high temperatures. For
$L\ll L_T^{*}$, we obtain from (\ref{eq:stiff.def}) and (\ref{eq:z.ferm})
\begin{eqnarray}
\label{eq:half.low}
&&\rho_s^{*} =\left\{
\begin{array}{l}
-( v^{*}/8{K^{*}}^2)(L_T^{*}/L),\;\;\mbox{for}\;N\;\mbox{even}\\
v^{*}/4\pi K^{*},\;\;\mbox{for}\;N\;\mbox{odd},
\end{array}
\right.
\end{eqnarray}
  while for $L\gg L_T^{*}$ we get
\begin{equation}
\label{eq:stiff.high}
 \rho_s^{*} =(-1)^{N+1}
2(L/L_T^{*})e^{-\pi \chi L/L_T^{*}},
\end{equation}
where $\chi=K^{*}+1/4K^{*}$. To obtain the value of $\rho_s$ away from
the fixed point, we
go back to the full action (\ref{eq:act.ferm}) and replace $(K_0, g_0)$
by $(K,g)$ from Eq.~(\ref{eq:marg}),
treating the deviations from the
fixed-point values as perturbations \cite{footnote3,RG,Affleck94}.
In first order, the Umklapp
term gives no contribution, while the perturbation in $K-K^{*}$ leads to:
\begin{eqnarray}
\label{eq:away}
\rho_s=\rho_s^{*}\times \left\{
\begin{array}{l}
1+1/[{\alpha}_N\ln(L/L_0)],\;\;\;\mbox{for}\;\;L\ll L_T^{*}\\
\exp\left(3L\pi/[8L_T^*\ln(T_0/T)]\right),\;\;\;\mbox{for}\;\;L\gg L_T^{*},
\end{array}\right.\end{eqnarray}
where $\alpha_N=1$ for even $N$, and $\alpha_N=2$ for odd $N$; the cut--off
$T_0$ is of the order of
$v^*/L_0$. The last equation is valid for $T\le T^{*}<T_0$, where
$L_T(T^{*})=a_0$. The $L$- and $T$-dependent corrections to $\rho_s^{*}$
resulting from the perturbation of the fixed-point action by the marginally
irrelevant operator are larger than those coming from the expansion of
Eq.~(\ref{eq:z.ferm}) in $(L/L_T^{*})^{\pm 1}$ at the fixed point.
In particular, the exponential
dependence of $\rho_s^{*}$ on $K^{*}$ in the high-temperature regime
results in a significant $T$-dependent
renormalization.
This renormalization may be
conjectured to remain  significant in the intermediate regime
$L\simeq L_T^{*}$ as well. We also note that the Umklapp processes lead
to the breakdown of the single-parameter scaling of $\rho_s$: the latter
scales with $L/L^{*}_T$ at the fixed-point but acquires
additional $L/ L_0$ and $T/T_0$ scalings away from the fixed point
(for a rough estimate, $L_0\simeq a_0$, and $T_0\simeq J_{ex}$).
This breakdown can be detected in numerical and
real experiments.
The fluctuations in $\rho_s$ can be calculated along the same lines as for
$\rho_s$ itself, and, in marked contrast to the integer $S$ case,
turn out to be exponentially small for
all $L/L_T$.

The negative value of $\rho_s$ for even $N$
simply reflects the fact that in this case the free energy has a maximum
at $\theta =0$, and an arbitrarily small twist drives
the system out of this state. Analysing
$\rho_s$ at finite $\theta$ (and low temperatures) we can see
 that $\rho_s$ vanishes at some $\theta^{*}\simeq L/L_T^{*}\ll1$ and then
 remains positive for all $\theta$, thus exhibiting a crossover from the
 paramagnetic to the diamagnetic regime. The parity effects in the spin
 stiffness are quite similar to that in persistent currents of
 electronic systems \cite{Byers,Legget,Loss}; in particular,
the result obtained above
can be checked without approximations for the special case of
the $XY$-model by mapping it on to the exactly solvable problem of free
fermions \cite{comm.par.}.

This work was supported by the  NSERC of Canada (D.L. and D.L.M.)
and in part by the US NSF through Grant No. DMR-89-20538 (D.L.M.).


\begin{references}
\bibitem{comm1} These results
have been verified up to two-loop order \cite{Zinn} and in
Monte-Carlo simulations \cite{Caffarel}.
\bibitem{comm.def.} Although $F$ is an even function in $\theta$, it is not
necessarily a function of $\theta^2$ (as we will see in the half-odd integer
spin case, $F$ can be a function of $|\theta|$). Thus, definition
(\ref{eq:stiff.def}) is more general than that used in previous works.
\bibitem{RG} In leading order, the flow of $v$ is slower than that of
$K$ and $g$, and we still can use the fixed point value $v^{*}$.
\end{references}
\end{document}